\documentclass[preprint,12pt]{elsarticle}

\usepackage{amssymb}
\usepackage{amsmath}
\usepackage{xcolor}

\usepackage{lineno}
\def\LG{\ensuremath{\mathcal{L}_{\mathrm{gauge}}}}
\def\LHY{\ensuremath{\mathcal{L}_{\mathrm{higgs+yukawa}}}}
\def\LFP{\ensuremath{\mathcal{L}_{\mathrm{fp}}}}
\def\LGF{\ensuremath{\mathcal{L}_{\mathrm{gf}}}}

\newcommand{\MS}{{\ensuremath{\overline{\mathrm{MS}}}}}

\DeclareMathOperator{\tr}{Tr}

\newcommand{\alam}{\ensuremath {a_\lambda}}
\newcommand{\as}{\ensuremath {a_s}}
\newcommand{\aw}{\ensuremath {a_2}}
\newcommand{\ay}{\ensuremath {a_1}}

\newcommand{\YtF}{\ensuremath {\mathcal{Y}_{f}}}
\newcommand{\YtFF}{\ensuremath {\mathcal{Y}_{ff'}}}
\newcommand{\YtFFF}{\ensuremath {\mathcal{Y}_{ff'f''}}}

\def\Yu{\ensuremath {\mathcal{Y}_{u}}}
\def\Yd{\ensuremath {\mathcal{Y}_{d}}}
\def\Yl{\ensuremath {\mathcal{Y}_{l}}}

\def\Yuu{\ensuremath {\mathcal{Y}_{uu}}}
\def\Yud{\ensuremath {\mathcal{Y}_{ud}}}
\def\Ydd{\ensuremath {\mathcal{Y}_{dd}}}
\def\Yll{\ensuremath {\mathcal{Y}_{ll}}}

\def\Yuuu{\ensuremath {\mathcal{Y}_{uuu}}}
\def\Yuud{\ensuremath {\mathcal{Y}_{uud}}}
\def\Yudd{\ensuremath {\mathcal{Y}_{udd}}}
\def\Yddd{\ensuremath {\mathcal{Y}_{ddd}}}
\def\Ylll{\ensuremath {\mathcal{Y}_{lll}}}

\newcommand{\witerms}[1]{\ensuremath{\textcolor{blue}{#1}}}

\journal{Physics Letters B}

\begin{document}

\begin{frontmatter}



\title{Three-loop anomalous dimensions \\ of fixed-charge operators in the SM}


\author{A.V.~Bednyakov}
\ead{bednya@jinr.ru}

\affiliation{organization={Joint Institute for Nuclear Research},
            addressline={Joliot-Curie, 6}, 
            city={Dubna},
            postcode={141980}, 
            state={Moscow region},
            country={Russia}}

\begin{abstract}
	In this Letter we consider renormalization of a class of scalar operators with fixed hypercharge $Q$ within the Standard Model. We carry out explicit computation of the corresponding anomalous dimensions up to the three-loop order. In spite of the fact that our result is gauge-dependent, in the Landau gauge and in the limit of vanishing weak isospin coupling the expression can be matched to recent gauge-independent computation based on the large-charge method. Our result serves an important and non-trivial cross-check of new developments in large-charge expansion and applications of the latter to realistic gauge theories. We not only confirm the leading and subleading terms in perturbative $Q$ expansion up to three loops, but also provide the expressions for sub-subleading coefficients that at the moment are not captured by the large-charge approach.   

\end{abstract}




\end{frontmatter}


\section{Introduction}
\label{sec:intro}

More than ten years have passed since three-loop renormalization-group (RG) equations  
\cite{Mihaila:2012fm,Bednyakov:2013cpa,Bednyakov:2014pia} for the full Standard Model (SM) became available in literature. These results raised immediate interest in scientific community due to Higgs boson discovery \cite{Aad:2012tfa,Chatrchyan:2012ufa} in 2012 and numerous discussions on vacuum (in)stability of the SM (see, e.g., Refs.~\cite{Bezrukov:2012sa,Degrassi:2012ry,Buttazzo:2013uya,Antipin:2013sga,Bednyakov:2015sca}). 

While during last decade many new results appeared that supplement and go beyond the three-loop approximation (see, e.g., Refs.~\cite{Bednyakov:2015ooa,Chetyrkin:2016ruf,Davies:2019onf}), in this paper we continue the study of the SM at this order. We carry out three-loop perturbative calculation of the anomalous dimensions for a class of local operators built from the Higgs field and having fixed weak hypercharge, which we denote here by $Q$. 
Such kind of computations play an important role in matching non-perturbative semiclassical results derived by  
large-charge expansion \cite{Hellerman:2015nra,Monin:2016jmo,Alvarez-Gaume:2016vff,Banerjee:2017fcx,Badel:2019oxl,Orlando:2019hte,Gaume:2020bmp,Antipin:2020abu,Giombi:2022gjj,Antipin:2022naw}
with perturbative calculations in theories with global symmetries (see, e.g., Refs.~\cite{Jack:2021ypd,Jack:2021lja,Jin:2022nqq,Bednyakov:2022guj}). 

Our results in the SM are gauge-dependent by construction and, obviously, do not have direct physical meaning. 
Nevertheless, recent application of the large-charge expansion to simple scalar electrodynamics in pioneering paper \cite{Antipin:2022hfe} demonstrated that perturbative results in the Landau gauge can be matched via state-operator correspondence to non-perturbative gauge-independent expressions. 

Despite the fact that phenomenological outcome of the large-charge ideas to realistic gauge theories of particle physics is far from obvious, the result of matching between non-perturbative (semi-classical) and perturbative calculations had important consequences to condensed-matter physics,i.e., singling out correct gauge-independent order parameter \cite{Antipin:2022hfe}.
This motivated us to go beyond scalar electrodynamics and consider the SM in its full glory. 

A natural question arises which symmetry of the SM can be used to define ``fixed-charge'' configurations and what operators correspond to the lowest-lying states in this sector. 
In our current study we does not address the first issue and refer to recent paper \cite{Antipin:2023tar} for details and discussions.
We naively consider unbroken SM and compute anomalous dimension of composite operators built from powers of (electrically) charged would-be goldstone boson field. The latter being a component of the Higgs doublet also posses weak hypercharge. 
Thus, we use powers of the charged would-be goldstone field as a proxy of operators with fixed weak hypercharge. More elaborate reasoning on this choice can be found in Ref~\cite{Antipin:2023tar}.  It is also important to  mention that in our computation we consider all gauge symmetries of the SM, i.e., $SU(3)_c \times SU(2)_W \times U(1)_Y$, and full flavour structure. 
However, matching of our expression with large-charge results for non-vanishing $SU(2)$ coupling is highly non-trivial, if possible at all, since non-Abelian nature of the latter requires to work with the phase where also gauge bosons condense \cite{Antipin:2023tar}, which breaks naive correspondence that we silently assume here.

The paper is organized as follows. In Section~\ref{sec:the_sm} we briefly review the SM Lagrangian with general Yukawa couplings. The definition of operators of our interest is given in Sec.~\ref{sec:operators} and some details of calculations are provided in Sec.~\ref{sec:details}. We present our results and conclusions in Sec.~\ref{sec:results_conclusions}.
\section{The Standard Model} 
\label{sec:the_sm}
We follow Ref.~\cite{Bednyakov:2012rb} and consider the Standard Model in the unbroken phase. The corresponding Lagrangian is given by
\begin{equation}
{\mathcal L} =
  \LG 
+ \LHY 
+ \LGF 
+ \LFP 
\label{eq:lag_classical}
\end{equation}
with 
\begin{align}
	\LG & = -\frac{1}{4} G^a_{\mu\nu} G^a_{\mu\nu}
	        -\frac{1}{4} W^i_{\mu\nu} W^i_{\mu\nu}
	        -\frac{1}{4} B_{\mu\nu} B_{\mu\nu} 
		+ (D_\mu \Phi)^\dagger (D_\mu \Phi)
		\nonumber\\
	    &
		+ i \sum\limits_j \left[\bar Q_j \hat D Q_j 
		+  \bar L_j \hat D L_j  
		+  \bar u_{j}^R \hat D u_{j}^R 
	+  \bar d_{j}^R \hat D d_{j}^R 
		+  \bar l_{j}^R \hat D l_{j}^R 
	\right]
		\label{eq:gauge}
\end{align}			
describing gauge interactions of gluons $G_\mu^a$, W-bosons $W_\mu^i$ and a B-boson $B_\mu$ with couplings $g_s$, $g_2$ and $g_1$, respectively. The left-handed quark $Q$ and lepton $L$ fields are $SU(2)_L$ doublets. Right-handed up-type $u^R$ and down-type $d^R$ quarks together with right-handed charged leptons $l^R$ are $SU(2)_L$ singlets and do not couple to W-bosons through covariant derivatives $\hat D$. The Higgs doublet has weak hypercharge $Y_W=1$ and can be decomposed as
\begin{align}
	\Phi =
	\left(
	\begin{array}{c}
		\phi^+(x) \\ \frac{1}{\sqrt 2} \left( h + i \chi \right)
		\end{array}
	\right),
	\qquad
	\Phi^c = i\sigma^2 \Phi^\dagger =
	\left(
	\begin{array}{c}
		\frac{1}{\sqrt 2} \left( h - i \chi \right) \\
		-\phi^-
		\end{array}
	\right)
	\label{eq:Phi_def}
\end{align}
with $\Phi^c$ being charge-conjugated doublet with $Y_W = -1$.
The self-interactions of $\Phi$ and Yukawa interactions of the latter with SM fermions are described by
\begin{align}
	\LHY & =  - \lambda \left( \Phi^\dagger \Phi \right)^2 
	-       \sum\limits_{i,j=1,2,3} \left[ Y^{ij}_l (L^L_i \Phi) l^R_j	
		+ \mathrm{h.c.}
	\right]
	\nonumber\\
	     & -  \sum\limits_{i,j=1,2,3} \left[
	Y^{ij}_u (Q^L_i \Phi^c) u^R_j
+ Y^{ij}_d (Q^L_i \Phi) d^R_j
+ \mathrm{h.c.} \right],
	\label{eq:ferm_higgs_lag}
\end{align}
where we emphasize the flavour structure of the SM described by matrix Yukawa couplings\footnote{In this computation we neglect possible contribution from right-handed neutrinos that can couple to the Higgs boson via the corresponding Yukawa coupling $Y_\nu$.} $Y_u$, $Y_d$, and $Y_l$. We omit the Higgs-boson mass parameter, since it does not affect our computation. As for the gauge-fixing $\LGF$ and ghost terms $\LFP$, we use here simple linear $R_\xi$ gauge:
\begin{align}
	\LGF  &=
	-\frac{1}{2\xi_\alpha} G_\alpha G_\alpha,
	      & \LFP &= - \bar c_\alpha \frac{\delta G_\alpha}{\delta \theta^\beta} c_\beta
	\label{eq:gauge_fix_fp_lag}
\end{align}
where $\alpha,\beta = (G,W,B)$, and $\delta G_\alpha/\delta \theta^\beta$ is the variation of the gauge-fixing function $G_\alpha = \partial_\mu A^\alpha_\mu$ [$A_\mu^\alpha = (G_\mu^a,W_\mu^i,B_\mu)$] 
under infinitesimal gauge transformations with parameters $\theta_\beta$ corresponding to $SU(3)_c \times SU(2)_L \times U(1)_Y$ local symmetries.

\section{Fixed-charge operators}
\label{sec:operators}
Let us now consider a family of local operators that have fixed hypercharge. We now denote it as $Q$ to be in contact with literature on large-charge expansion. We naively transfer our experience in theories with global \cite{Bednyakov:2022guj} and local symmetries \cite{Antipin:2022hfe} and study the renormalization of $\mathcal{O}_Q$ built from would-be goldstone boson fields $\phi^+$
\begin{align}
	\mathcal{O}_Q = (\phi^+)^Q  \equiv \underbrace{\Phi^1\ldots \Phi^1}_{Q},
	\label{eq:operators}
\end{align}
where we indicate that the considered operators at fixed $Q$ belong to $(Q+1)$-plet that transforms under $SU(2)_L$. Due to this symmetry, all components of the multiplets renormalize in the same way and have the same anomalous dimensions. 
In what follows we silently assume dimensional regularization and define all the renormalization constants in the (modified) minimal-subtraction scheme $\MS$. 

One can use simple combinatorial argument and demonstrate  that the anomalous dimension $\gamma_Q$ of operators \eqref{eq:operators} for general $Q$ can be written in perturbation theory as
\begin{align}
	\gamma_Q = \sum\limits_{l=1}^{\infty} \gamma_Q^{(l-\mathrm{loop})}, \qquad
	\gamma_Q^{(l-\mathrm{loop})} \equiv \sum\limits_{k=0}^{l} C_{lk} Q^{l+1 -k},
	\label{eq:general_adm_Q_dep}
\end{align}
where the coefficients $C_{lk}$ depend on the SM couplings and gauge-fixing parameters.
One sees that $l$-loop contribution is a polynomial of degree $l+1$ in $Q$. As a consequence, at three loops we can recover full dependence of $\gamma^{(3)}_Q$ on $Q$ by 
carrying out explicit computations for fixed $Q=1,2,3,4$.  

The case $Q=1$ corresponds to the Higgs field itself. The renormalization of the latter is known from literature and it is not a problem to derive the  anomalous dimension from the relation between bare $\Phi_0$ and renormalized $\Phi$ fields:
\begin{align}
	\Phi_0 = Z^{1/2}_\Phi \Phi, \qquad \gamma_\Phi \equiv \frac{d Z_\Phi}{ d \ln \mu}\cdot Z^{-1}_\Phi = \gamma_{Q=1}.
\end{align}
Here $\mu$ denotes the renormalization scale. 
The cases $Q=2,3,4$ were considered separately. We computed three-loop Green functions with $\mathcal{O}_Q$ insertions and utilized the relation 
\begin{align}
	[\mathcal{O}_Q]_R = Z_{Q} [\mathcal{O}_Q], \qquad \gamma_Q \equiv 
	-  \frac{d Z_Q}{ d \ln \mu}\cdot Z^{-1}_Q 
	\label{eq:ZQ_to_gQ}
\end{align}
to derive the corresponding anomalous dimensions $\gamma_{Q=2,3,4}$. 

\label{sec:details}
\section{Details of Calculation}
Let us briefly review the main steps of our calculation that heavily relies on the computer setup which we used to compute three-loop RG functions in the SM  with matrix Yukawa couplings \cite{Bednyakov:2013cpa,Bednyakov:2014pia}. We minimally modified the SM model file for the \texttt{DIANA} \cite{Tentyukov:1999is} package\footnote{based on \texttt{QGRAF} \cite{Nogueira:1991ex}} to generate relevant Feynman graphs with $(\phi^+)^n$ operator insertions into Green functions having  $n$ external $\phi^+$ legs. 
We considered only $n=2,3,4$ cases and extracted the corresponding renormalization constants $Z_{Q=2,3,4}$ from the ultraviolet (UV) divergences of one-particle-irreducible bare Green functions $\Gamma^{\mathrm{bare}}_{(\phi^+)^n}$. We required that the renormalized counterpart $\Gamma^{\mathrm{R}}_{(\phi^+)^n}$ 
\begin{align}
	\Gamma^{\mathrm{R}}_{(\phi^+)^n}(q, p_1, ..., p_n) = Z_{Q=n} Z_\Phi^{\frac{n}{2}} \cdot \Gamma^{\mathrm{bare}}_{(\phi^+)^n}(q, p_1, ..., p_n), \quad q = \sum\limits_{i=1}^n p_i,
\end{align}
should be finite. To compute the UV part of the Green functions, we set all external momenta to zero $p_i=0$ and introduced an auxiliary mass into each internal propagator.
This infra-red rearrangement trick pioneered in Ref.~\cite{Vladimirov:1979zm} and elaborated in Ref.~\cite{Misiak:1994zw,Chetyrkin:1997fm} avoids spurious infrared divergences but requires additional counterterms for the SM boson masses. The approach gives rise to single-scale vacuum integrals that can be easily evaluated by means of the \texttt{MATAD} code \cite{Steinhauser:2000ry} written in \texttt{FORM} \cite{Kuipers:2012rf}.  Let us finally mention that we routinely used \texttt{COLOR} \cite{vanRitbergen:1998pn} to compute $SU(3)_c$ group factors and applied (semi-)naive treatment \cite{Chetyrkin:2012rz} of the $\gamma_5$ matrix in dimensional regularization.  

\section{Results and Conclusions}
\label{sec:results_conclusions}
Having at hand $\gamma_Q$ for fixed $Q=1,2,3,4$, one can derive the expression for all the coefficients $C_{lk}$ up to $l=3$. In what follows we present our result in the Landau gauge $\xi_\alpha = 0$ and use the following notation for gauge couplings and Higgs self-interaction
\begin{align}
	a_s = \frac{g_s^2}{16\pi^2}, \quad
	a_1 = \frac{g_1^2}{16 \pi^2}, \quad
	a_2 = \frac{g_2^2}{16\pi^2}, \quad
	a_\lambda = \frac{\lambda}{16\pi^2}. 
	\label{eq:couplants}
\end{align}
For the traces of Yukawa matrices that appear in the result we introduce the abbreviations
\begin{align}
	\YtF & =  \frac{\tr Y_f Y_f^\dagger}{16\pi^2},\quad 
	\YtFF =  \frac{\tr Y_f Y_f^\dagger Y_{f'} Y_{f'}^\dagger}{(16\pi^2)^2},  \nonumber\\
	\YtFFF & =  \frac{\tr Y_f Y_f^\dagger 
	Y_{f'} Y_{f'}^\dagger Y_{f''} Y_{f''}^\dagger
}{(16\pi^2)^3}, 
	\quad f,f',f''\in \{u, d, l\}.
	\label{eq:yukawa_traces}
\end{align}
The leading coefficients in $Q$ expansion up to three loops are given by\footnote{Note that we should rescale $\lambda \to \lambda/6$ to make direct comparison with Ref.~\cite{Antipin:2023tar}.}
\begin{align}
	C_{10} & =  2 \alam, & C_{20} & =  -8 \alam^2, & C_{30} & =  64 \alam^3
    \label{eq:leading_Q_3l}
\end{align}
and coincide with those of $\phi^4$ theory. 
The expressions for subleading coefficients take the form 
\begin{align}
	C_{11}  = & -\frac{3 \ay}{4}-2 \alam+3 \Yd +3 \Yu+\Yl \witerms{-\frac{9 \aw}{4}}, 
    \label{eq:subleading_Q_1l}
	\\
	C_{21}  = & \frac{\ay^2}{16}-\ay \left(2 \alam - \witerms{\frac{\aw}{8}}\right)+\witerms{\frac{3 \aw^2}{16}} \nonumber \\
	       & +4 \alam^2-4 \alam (3 \Yu + 3 \Yd + \Yl)+ 2 (3 \Yuu + 3 \Ydd +\Yll), 
    \label{eq:subleading_Q_2l}
	       \\
C_{31}  = & -\frac{1}{16} \left(\ay^3 + \witerms{3 \ay^2 \aw 
	+ 3 \ay \aw^2 + 3 \aw^3}  \right)  \left(1 - 9 \zeta_3\right) -\alam \left[\ay (\ay + \witerms{2 \aw} ) (1 + 3 \zeta_3) \right] \nonumber \\
       & + 12 \alam^2 \ay(3-2 \zeta_3) 
	-32 \alam^3 (8 - 9\zeta_3) 
       +16 \alam^2 (3 \Yu+ 3 \Yd + \Yl)
       \nonumber \\
       & 
       -8 \alam (3 \Yuu + 3 \Ydd + \Yll) (1 - 3\zeta_3)       
       -8\zeta_3 \left(3 \Yuuu + 3 \Yddd + \Ylll\right) 
       \nonumber\\
       &
       + \witerms{4 \alam^2 \aw (13 -12 \zeta_3)}          
       - \witerms{\alam \aw^2 (5 + 12 \zeta_3)},
    \label{eq:subleading_Q_3l}
	\end{align}
	where we highlight terms proportional to the $SU(2)_L$ coupling $g_2$.
	Again contributions involving only $\alam$ reproduce the expressions in $O(4)$-symmetric $\phi^4$ model. 

	All the above-mentioned coefficients can be compared to all-order results \cite{Antipin:2023tar} in the limit $g_2 = 0$. We go beyond this approximation and provide also subsubleading terms. At two loops we have 
\begin{align}
C_{22} & =  \frac{425 \ay^2}{96}-\witerms{\frac{277 \aw^2}{32}} +  \ay \left[2 \alam+\frac{5}{24} (5 \Yd+15 \Yl+17 \Yu) + \witerms{\frac{7 \aw}{16}} \right] \nonumber\\
       &+\witerms{\frac{15}{8} \aw (3 \Yd+\Yl+3 \Yu)} +10 \alam^2+4 \alam (3 \Yd+\Yl+3 \Yu) \nonumber \\
       & +20 \as (\Yd+\Yu)-\frac{1}{4} (51 \Ydd + 51 \Yuu + 17 \Yll-6 \Yud). 
\end{align}
At three loops we extract two coefficients that are not predicted by semi-classical approach. To save space\footnote{Full results in the linear $R_\xi$ gauge  are available as supplementary \texttt{Mathematica} files.} we present them in the limit  $g_2=0$:
{\allowdisplaybreaks
\begin{align}
C_{32} & =  \frac{9}{16} \ay^3 (14 - 3 \zeta_3) +\frac{3}{32} \alam \ay^2 (16 \zeta_3-267)
-4 \ay \alam^2 (32 - 33 \zeta_3)
\nonumber\\
       & - \frac{\ay^2}{6} \left[\Yd (13 + 3\zeta_3)- \Yl (69 - 81 \zeta_3)-\Yu (53 - 75 \zeta_3)\right] 
       \nonumber\\
             & 
	     +\frac{\ay \alam}{8} \left[\Yd (179 + 224 \zeta_3)- \Yl (167 - 288 \zeta_3) - \Yu (25 - 416 \zeta_3)\right] 
	     \nonumber\\
       & +\frac{\ay}{6} \left[\Ydd (51 \zeta_3-146)+ \Yll (18 -63 \zeta_3)- \Yuu (86 + 21\zeta_3)\right] \nonumber\\
       & + 2 \alam^3 (353 - 432 \zeta_3)-8 \alam^2 (3 \Yu + 3 \Yd+\Yl) 
       \nonumber\\
       & -12 \alam \as (17 - 16 \zeta_3) (\Yd+\Yu)+16 \as (5-6 \zeta_3) (\Ydd+\Yuu) 
       \nonumber\\
       & - 6 \alam \left[
	9 \Yd^2 + 9\Yu^2 + \Yl^2
	+6 (\Yu + \Yd) \Yl
	+18 \Yd \Yu
\right] \nonumber\\
       & 
+ \frac{\alam}{4}
\left[
	(3 \Yuu + 3 \Ydd + \Yll) (235 - 240 \zeta_3)
	+\Yud ( 510 - 576 \zeta_3)
\right] \nonumber\\
       & +4 \left[
       9  (\Yd + \Yu)  (\Ydd + \Yuu)
       +3 
       (\Yd + \Yu) \Yll
       + 3 (\Ydd + \Yuu) \Yl
       + \Yl \Yll
\right] \nonumber\\
       &
       +6 (\Yudd + \Yuud)
       -2(3 \Yddd + 3 \Yuuu + \Ylll) (11 - 12 \zeta_3)
       ,
       \label{eq:C32}
       \\
%
C_{33} & =  \frac{\ay^3}{432} (23678-13275 \zeta_3)
	+\frac{\ay^2 \alam}{32} (911-96 \zeta_3)
+\ay \alam^2 (107-108 \zeta_3)
\nonumber\\
       &
-\frac{\ay^2}{3456}\left[174799 \Yu + 20311 \Yd+200637 \Yl - 288 \zeta_3 (151 \Yu-23 \Yd+279 \Yl)\right] \nonumber\\
       & 
       +\frac{\ay \alam}{8} [25 \Yu -179 \Yd+167 \Yl -32 \zeta_3 (7 \Yd+9 \Yl+13 \Yu)]
       \nonumber\\
       &
+\frac{11}{4} \ay^2 \as (15 - 16 \zeta_3)
       +\frac{\ay \as}{36} [\Yd (720 \zeta_3-991)+\Yu (2448 \zeta_3-2419)] \nonumber\\
       & + \left( 6 \alam -\frac{3\ay }{4}\right)(3 \Yu + 3 \Yd +\Yl)^2 
       -\frac{61\alam^2}{2} (3 \Yu + 3 \Yd + \Yl)
       \nonumber\\
       & +\frac{\ay}{48} [55 \Yuu + 259 \Ydd-243 \Yll-186 \Yud + 24 \zeta_3 (\Yuu + \Ydd +3 \Yll+16 \Yud)] \nonumber\\
       &
       +\frac{\alam}{4} (144 \zeta_3 (3 \Yuu + 3 \Ydd+\Yll+4 \Yud)
       -143 (3 \Yuu + 3 \Ydd + \Yll) -510 \Yud) \nonumber\\
       &+\frac{2 \as^2}{3} (311-36 \zeta_3) (\Yd+\Yu) 
       -12 \alam \as (16 \zeta_3-17) (\Yd+\Yu) 
       \nonumber\\
       & -\frac{\as}{2} [145 \Yuu + 145 \Ydd -114 \Yud -48 \zeta_3 (\Yuu + \Ydd -2 \Yud)] \nonumber \\
       & +\frac{1}{2}  (12 \Yuu + 12 \Ydd +4 \Yll-3 \Yud)(3 \Yu+ 3 \Yd + \Yl) 
	-2 \alam^3 (275 - 288 \zeta_3) \nonumber\\
       & + \frac{1}{16} \left[(327 - 208 \zeta_3) ( 3 \Yuuu + 3 \Yddd  + \Ylll)-57 (\Yudd+\Yuud)\right].
       \label{eq:C33}
\end{align}
}

Let us make a few comments on the obtained result.
First of all, we stress that it is \emph{not} gauge-independent. Secondly, one can notice that there is no dependence on the strong coupling $\as$ in first two leading coefficient $C_{l0}$ and $C_{l1}$,  meaning that strong interactions play a role of ``spectator'' when subsubleading terms are ignored. Finally, $C_{l0}$ and $C_{l1}$, when evaluated in the Landau gauge and in the limit of vanishing $g_2$, can be compared to large-charge all-order predictions of recent study \cite{Antipin:2023tar}. Perfect agreement that was found serves as a welcome cross-check of both computations. 

It is also worth mentioning that the anomalous dimensions obtained in Ref.~\cite{Antipin:2023tar} are gauge-independent by construction. As a consequence, they should correspond to some lowest-lying gauge-independent operators with fixed hypercharge and can not be directly connected to our current result based on Eq.~\eqref{eq:operators}. Nevertheless, when non-Abelian interactions are ignored, one can follow the reasoning of \cite{Antipin:2022hfe} and introduce a generalization of our operators $\mathcal{O}_Q$ that accounts for dressing of the charge with coherent state of Abelian gauge bosons describing Coulomb-like field. In the Landau gauge these gauge-independent operators can be reduced to $\mathcal{O}_Q$ given in Eq.~\eqref{eq:operators}, thus, explaining the coincidence.

For non-Abelian symmetries the trick seems not to be applicable in the naive form, and the case $g_2\neq 0$ requires further investigation \cite{Antipin:2023tar}. However, we believe that our present result can shed light on this issue. We also note that the Higgs boson observed at the LHC, although being electrically neutral, also posses weak hypercharge. As a consequence, the properties of operators with large hypercharge can be important, e.g., in connection with multiple higgs production  (see, e.g., recent discussion \cite{Demidov:2022ljh} and references therein).

\section{Acknowledgments}
\label{sec:acknowledgement}
We appreciate fruitful discussions with Oleg Antipin on the application of large-charge expansion in gauge theories. We also grateful to Francesco Sannino for his comments on the manuscript and thank all the authors of Ref.~\cite{Antipin:2023tar} for sharing with us the preliminary results of their inspiring paper.
\bibliographystyle{elsarticle-num.bst}
\bibliography{references.bib}

\end{document}